\documentclass[a4paper,11pt]{article}
\pdfoutput=1 

\usepackage{jcappub} 

\usepackage[T1]{fontenc} 
\usepackage{color}
\usepackage{textcomp}

\usepackage{soul}
\usepackage{ulem}

\usepackage{import}

\usepackage{comment}

\title{Inflationary Initial Conditions for the Cosmological Gravitational Wave Background}

\author[a,b]{Lorenzo Valbusa Dall'Armi,}
\author[a,b]{Alina Mierna,}
\author[a,b,c,d]{Sabino Matarrese,}
\author[e,f,a]{Angelo Ricciardone}

\affiliation[a]{Dipartimento di Fisica e Astronomia ``G. Galilei'',
Universit\`a degli Studi di Padova, via Marzolo 8, I-35131 Padova, Italy}
\affiliation[b]{INFN, Sezione di Padova,
via Marzolo 8, I-35131 Padova, Italy}
\affiliation[c]{INAF- Osservatorio Astronomico di Padova, \\ Vicolo dell’Osservatorio 5, I-35122 Padova, Italy}
\affiliation[d]{Gran Sasso Science Institute, Viale F. Crispi 7, I-67100 L’Aquila, Italy}
\affiliation[e]{Dipartimento di Fisica “Enrico Fermi”, Universit\`a di Pisa, Pisa I-56127, Italy}
\affiliation[f]{INFN sezione di Pisa, Pisa I-56127, Italy}

\begin{document}

\abstract{
The initial conditions on the anisotropies of the stochastic gravitational-wave background of cosmological origin (CGWB) largely depend on the mechanism that generates the gravitational waves. Since the CGWB is expected to be non-thermal, the computation of the initial conditions could be more challenging w.r.t. the Cosmic Microwave Background (CMB), whose interactions with other particles in the early Universe lead to a blackbody spectrum. In this paper, we show that the initial conditions for the cosmological background
generated by quantum fluctuations of the metric during inflation deviate from adiabaticity. These primordial gravitational waves are indeed generated by quantum fluctuations of two independent degrees of freedom (the two polarization states of the gravitons). Furthermore, the CGWB plays a negligible role in the Einstein's equations, because its energy density is subdominant w.r.t. ordinary matter. Therefore, the only possible way to compute the initial conditions for inflationary gravitons is to perturb the energy-momentum tensor of the gravitational field defined in terms of the small-scale tensor perturbation of the metric. This new and self-consistent approach shows that a large, non-adiabatic initial condition is present even during the single-field inflation. Such a contribution enhances the total angular power spectrum of the CGWB compared to the standard adiabatic case, increasing also the sensitivity of the anisotropies to the presence of relativistic and decoupled particles in the early Universe. In this work we have also proved that our findings are quite general and apply to both single-field inflation and other scenarios in which the CGWB is generated by the quantum fluctuations of the metric, like the curvaton. 
}

\maketitle

\section{Introduction}

The detection of a stochastic gravitational wave background (SGWB) by the PTA collaboration (i.e., NANOGrav, EPTA/InPTA, PPTA, and CPTA) opens a new window for the study of early Universe cosmology~\cite{NANOGrav:2023gor, EPTA:2023fyk, Reardon:2023gzh, Xu:2023wog}. The analysis of the possible inflationary and astrophysical interpretations of the detected signal has been done in~\cite{NANOGrav:2023hvm, NANOGrav:2023hfp, EPTA:2023xxk, Franciolini:2023pbf,Ellis:2023tsl,Franciolini:2023wjm,Vagnozzi:2023lwo,Figueroa:2023zhu,Liu:2023ymk}. The current ground-based interferometers, such as LIGO, Virgo and KAGRA, are close to reaching the sensitivity to detect the astrophysical gravitational wave background (AGWB) from
unresolved sources, while 
future space-based, such as LISA~\cite{LISA:2017pwj}, Taiji~\cite{Orlando:2020oko}, BBO~\cite{Corbin:2005ny} and DECIGO~\cite{Kawamura:2006up}, and ground-based interferometers, such as Einstein Telescope~\cite{Punturo:2010zz, Maggiore:2019uih, Branchesi:2023mws} and Cosmic Explorer~\cite{Reitze:2019iox, LIGOScientific:2016wof}, might be able to detect also the cosmological gravitational wave background (CGWB) produced by various mechanisms in the early Universe, such as inflation, phase transitions, cosmic strings~\cite{Guzzetti:2016mkm,Bartolo:2016ami,Caprini:2018mtu}. Therefore, it is necessary to find a method to differentiate the cosmological GW signal from the astrophysical one. The frequency dependence is one of the possible ways to distinguish among the various GW backgrounds~\cite{Caprini:2019pxz, Flauger:2020qyi}, since the vast majority of the proposed cosmological sources produce spectra which peak at some characteristic scales or deviate from the $f^{2/3}$ power-law scaling~\cite{Phinney:2001di} predicted for the astrophysical background generated by binary systems during the inspiral stage. However, because of the better angular resolution of future interferometers, anisotropies in GW energy
density could provide a new tool to distinguish among various sources of GWs in the early Universe~\cite{Bartolo:2019oiq,Bartolo:2019yeu}, see e.g.~\cite{Cui:2023dlo,Contaldi:2023jcc} for recent applications.

To characterize the anisotropies of the CGWB, similarly to CMB~\cite{Dodelson:2003ft}, we can use the Boltzmann equation for the graviton distribution function ~\cite{Contaldi:2016koz, Bartolo:2019oiq, Bartolo:2019yeu}. The CMB anisotropies are generated only at the last scattering surface, since any prior information is erased by the collisions photons suffer before recombination. This is in contrast with the CGWB anisotropies, which propagate freely at all energies below the Planck scale providing unique information about the primordial Universe. Anisotropies in GW energy density are induced by the GW production mechanism (initial conditions) and by the GW propagation through the large scale metric perturbations of the Universe (Sachs-Wolfe and Integrated Sachs-Wolfe effects). The initial anisotropies are greatly dependent on the properties of the GW source.

In the case of adiabatic initial conditions, the GW energy density perturbations at early times are equal to those of the CMB photons. When more than one scalar field is present during inflation, non-adiabatic modes could be generated~\cite{Hu:1996yt}. If the CGWB is the product of the inflaton decay during reheating along with photons and baryons, the initial conditions are adiabatic if no other fields contribute to the curvature perturbation, since the inhomogeneities of the source are inherited by the decay products. This is compatible with the ‘‘Separate Universe'' approach~\cite{Wands:2000dp}, which argues that each Hubble patch evolves like a separate Robertson–Walker Universe. Different regions experience indeed the same evolution along a single phase-space trajectory, separated only by a shift in the expansion. When a single-clock mechanism sets the initial curvature perturbations of the Universe, the perturbations must be adiabatic.
Conversely, if the GW production involves a local time-shifting function, different regions are not simply time translations of each other and thus non-adiabatic modes could appear. 

In this paper, we focus our analysis on the CGWB produced by the intrinsic quantum fluctuations of the metric during inflation. In this scenario, the initial overdensity of the CGWB is not related to the perturbation of the inflaton, since it arises from two additional independent degrees of freedom - the two polarizations of the tensor perturbations. The quantum fluctuations of the metric would be present indeed also if there is no clock (pure de Sitter). Therefore, the standard adiabatic initial condition is no longer valid. Furthermore, the Einstein equations are blind to the presence of the CGWB, because its energy density is subdominant w.r.t. the standard radiation. This implies that, in order to obtain the CGWB energy density perturbation at early times one has to compute explicitly the energy-momentum tensor of GWs. The energy-momentum tensor of GWs can be computed in terms of covariant derivatives of the radiative degrees of freedom of the metric~\cite{Isaacson:1967sln} or by perturbing the Einstein tensor~\cite{Landau:1975pou}. In both cases, one should take into account terms quadratic in the GWs of high-frequency, according to the shortwave approximation, and for terms of order zero and linear in the large-scale perturbations of the Universe. This method is analogous to the computation of isocurvature fluctuations during inflation~\cite{Mollerach:1989hu}. Note also that this approach, which allows to use a microscopic quantity (i.e., $h_{ij}$) to have a macroscopic description (i.e., $\rho$) of the system, is quite general and could also be applied to the perturbations induced by spectator (scalar) fields that fluctuate during inflation independently on the inflaton.  

The non-adiabatic initial condition computed in this work is approximately equal to two times the adiabatic initial condition, with opposite sign. This difference enhances the total angular power spectrum of the CGWB background by an order of magnitude, having important implications for the detectability of the anisotropies. In particular, a large initial condition from inflation could be used to overcome the sample variance limit in the anisotropies due to fluctuations of the monopole at interferometers discussed in~\cite{Mentasti:2023icu,Mentasti:2023gmg}.
In addition, the non-adiabatic initial condition enhances the sensitivity of the angular power-spectrum of the CGWB to the abundance of extra relativistic degrees of freedom at early times~\cite{ValbusaDallArmi:2020ifo}. Moreover, the cross-correlation between the CGWB and the Cosmic Microwave Background (CMB) anisotropies~\cite{Ricciardone:2021kel, Braglia:2021fxn} could decrease in presence of non-adiabatic initial conditions. 
 
The paper is organized as follows. In Section \ref{Boltzmann equation for GWs} we outline the formalism for describing the CGWB
anisotropies. In Section \ref{Adiabatic initial conditions for the CGWB}
we consider an example of adiabatic initial conditions for the CGWB to understand the similarities and differences w.r.t. the CMB. In Section \ref{Non-adiabaticity of gravitons during inflation} we show that the GW induced by the quantum fluctuations of the metric are non-adiabatic. 
In Section \ref{Inflationary initial conditions for the CGWB} we introduce the new approach for the computation of the initial graviton overdensity from inflation. In Section~\ref{sec:The angular power spectrum of the CGWB} we compute the scalar contribution to the angular power spectrum of the CGWB and we study its dependence for non-adiabatic initial conditions on decoupled and relativistic species. In Section \ref{Tensor contribution to the angular power spectrum} we discuss how the non-adiabatic initial conditions change the tensor contribution to the angular power spectrum of the CGWB. In Section \ref{Cross-correlation between the CGWB and the CMB} we analyze the correlation between the CGWB and the CMB anisotropies. In Section \ref{An example: Curvaton mechanism} we show that the initial conditions derived in this work are valid also for primordial GWs in the curvaton scenario. 

In the present work we provide a full derivation and an extension of the results anticipated in~\cite{ValbusaDallArmi:2023nqn}.

 \section{Boltzmann equation for GWs}
 \label{Boltzmann equation for GWs}
 
We consider a perturbed Friedmann-Lemaître-Robertson-Walker (FLRW) metric in the Poisson gauge 
\begin{equation}
ds^2 = a^2\left[-e^{2\Psi}d\eta^2 +e^{-2\Phi}\left(e^{\gamma}\right)_{ij}dx^idx^j\right],
\end{equation}
where  $\Phi$ and $\Psi$ represent the large-scale scalar perturbations and $\gamma_{ij}$ is the transverse-traceless tensor perturbation. We identify GWs as small-scale tensor modes with comoving momentum $q$ much smaller than the typical scales $k$ over which the background varies. Since we consider perturbations of scales separated by several orders of magnitude, it is possible to split the tensor perturbations $\gamma_{ij}$ into the small-scale ripples $h_{ij}$ and the large-scale tensor modes $H_{ij}$, $\gamma_{ij}\equiv h_{ij}+H_{ij}\,$. In this case, we can use the shortwave approximation~\cite{Isaacson:1967sln} to describe the propagation of GWs on a curved background. We introduce then a graviton
distribution function $f_{\rm GW} = f_{\rm GW}(x^{\mu}, p^{\mu})$, which is function of position $x^{\mu}$ and momentum $p^{\mu} = dx^{\mu}/d\lambda$, where $\lambda$ is an affine parameter along the GW trajectory. The evolution of the CGWB in the phase space is therefore described by the Boltzmann equation ~\cite{Contaldi:2016koz,Bartolo:2019oiq,Bartolo:2019yeu}. The Boltzmann equation for the graviton distribution function $f_{\rm GW}$ is
\begin{equation}
 \mathcal{L}\left[f_{\rm GW}\right] = \mathcal{C}\left[f_{\rm GW}\right] + \mathcal{I}\left[f_{\rm GW}\right] \, , 
\end{equation}
where $\mathcal{L}\left[f\right] = d f/d\lambda $ is the Liouville operator, $\mathcal{C}\left[f\right]$ accounts for the collision of GWs and $\mathcal{I}\left[f\right]$ for their emissivity from cosmological and astrophysical sources. The collision of GWs has an impact on the graviton distribution only at higher order and therefore at the first order in perturbations it can be disregarded~\cite{Misner:1973prb,Bartolo:2018igk}. For the cosmological sources, the emissivity term can be regarded as an initial condition on the GW distribution. Solving the Boltzmann equation at first order around a FLRW metric leads to
 \begin{equation}
  \frac{\partial f_{\rm GW} }{\partial \eta} + n^i\frac{\partial f_{\rm GW} }{\partial x^i} + \left(\frac{\partial \Phi}{\partial \eta} - n^i\frac{\partial\Psi}{\partial x^i} - \frac{1}{2}\frac{\partial H_{jk}}{\partial \eta}n^jn^k\right)q\frac{\partial f_{\rm GW} }{\partial q} = 0,
 \end{equation}
 where $n^i$ is the direction along the GW trajectory. The first two terms represent a free streaming term, namely the propagation of GWs at all scales. The third term encodes the redshifting of gravitons, including the Sachs-Wolfe (SW) and the Integrated Sachs-Wolfe (ISW) effects. The graviton distribution function $f_{\rm GW} $ can be expanded as the isotropic component $\bar{f}_{\rm GW} $ that solves the Boltzmann equation at zero order and is sensitive only to the expansion of the Universe and an anisotropic contribution that solves the Boltzmann equation at first order and can be parameterized in terms of the dimensionless function $\Gamma$ as
\begin{equation}
    \delta f_{\rm GW}(\eta,\vec{x},\hat{n}, q) = -q\, \frac{\partial\bar{f}_{\rm GW}}{\partial q}(q) \, \Gamma(\eta,\Vec{x}, \hat{n}, q) \, .
\end{equation}
The energy density of the CGWB can be written in terms of the graviton distribution function at a given time and location as
\begin{equation}
    \rho_{\rm GW} (\eta, \Vec{x})= \frac{1}{a^4\left(\eta\right)}\int \frac{d^3q}{(2\pi)^3}  \, q \, f_{\rm GW}(\eta,\vec{x},\hat{n}, q) \,  =  \rho_{c } \int d \, \ln \, q \int d\hat{n}\,   \Omega_{\rm GW}(\eta,\vec{x},\hat{n}, q) \, . 
\end{equation}
We can introduce the energy density contrast of the CGWB defined as
   \begin{equation}
     \delta_{\rm{GW}} (\eta_0,\Vec{x}_0, \hat{n}, q) \equiv \frac{\Omega_{\rm{GW}}(\eta_0,\Vec{x}_0, \hat{n}, q)  -  \bar{\Omega}_{\rm GW}\left(\eta_0,q\right)}{ \bar{\Omega}_{\rm GW}\left(\eta_0,q\right)}\,,
    \end{equation}
where $x_0$ denotes the location of the observer, which could be taken as the origin, $\eta_{\rm 0}$ is the present time and the quantity $ \bar{\Omega}_{\rm GW}\left(\eta_0,q\right)$  is defined as the spatial average. The energy density contrast is related to the function $\Gamma$ and the spectral energy density of GWs through
    \begin{equation}
     \delta_{\rm GW} (\eta_0,\Vec{x}_0, \hat{n}, q)= \left[ 4 -  n_{\rm gwb}(q)\right]\Gamma(\eta_0,\Vec{x}_0, \hat{n}, q) \,,
 \end{equation}
where $n_{\rm gwb}$ is the Gravitational Wave Background (GWB) spectral index
\begin{equation}
    n_{\rm gwb}(q) \equiv  \frac{ \partial \ln \, \bar{\Omega}_{\rm{GW}}\left(\eta_0, q\right)}{\partial  \ln  q}.
 \end{equation}
The solution of the Boltzmann equation in Fourier space is
\begin{equation}
    \begin{split}
        \Gamma(\eta_0,\vec{k},\hat{n},q) = \Gamma(\eta_{\rm in},\vec{k},\hat{n},q)e^{ik\mu(\eta_{\rm in}-\eta_0)}&+\Psi(\eta_{\rm in},\vec{k})e^{ik\mu(\eta_{\rm in}-\eta_0)}-\Psi(\eta_0,\vec{k})\\
        &+\int_{\eta_{\rm in}}^{\eta_0} d \tilde{\eta} \left[\Phi^\prime(\tilde{\eta},\vec{k})+\Psi^\prime(\tilde{\eta},\vec{k})-\frac{1}{2}\hat{n}^i \hat{n}^j H_{ij}^\prime(\tilde{\eta},\vec{k})\right]e^{ik\mu\tilde{\eta}} \, , 
        \label{eq:Gamma0_solution}
    \end{split}
\end{equation}
where $\mu\equiv \hat{k}\cdot \hat{n}$. The first term is due to the initial conditions, while the other ones are related to the propagation of the GWs through the large-scale scalar and tensor perturbations of the Universe. Assuming statistical isotropy, the angular power-spectrum of the CGWB is defined as
\begin{equation}
\label{power-spectrum}
    \begin{split}
        \langle \delta_{{\rm GW},\ell m}\delta^*_{{\rm GW},\ell'm'} \rangle = \delta_{\ell \ell'}\delta_{mm'}C_\ell \, .
    \end{split}
\end{equation}
In this work, we focus on the initial condition contribution to the angular power spectrum, computing explicitly the perturbation of the distribution function of the CGWB at the production, $\Gamma(\eta_{\rm in},\vec{k},\hat{n},q)$.

\section{Adiabatic initial conditions for the CGWB}
\label{Adiabatic initial conditions for the CGWB}

We start by illustrating the example in which the perturbations of the CGWB are adiabatic, in order to understand the analogies and the differences w.r.t. the CMB. Apart from the quantum fluctuations of the metric, a CGWB could be produced by the decay of the inflaton during reheating, along with other particle species that fill the universe. In this case, the initial conditions for the energy density of the CGWB are adiabatic, because they are connected to the perturbations of the inflaton, since the inhomogeneities in the inflaton field propagate to its decay products. When the inflaton begins to oscillate around the minimum of its potential and decay into the standard radiation and gravitons, the energy conservation equation perturbed at first order for each fluid in the large scale limit reads~\cite{Malik:2002jb, Weinberg:2003sw, Weinberg:2004kr}
\begin{equation}
\label{con. Eq.}
    \delta\rho^\prime_i + 3\mathcal{H}(1+w_i)\delta\rho_i - 3\Phi^\prime(1+w_i)\rho_i = a\left( Q_i\Psi + \delta Q_i\right) \, , 
\end{equation}
where $\mathcal{H}$ is the conformal Hubble parameter, $Q_{i}\,$ is the local energy-momentum transfer of each component $(i = \rm{rad}, \rm{GW}, \rm{\varphi})$, which is constrained by $\sum_iQ_{i} = 0\,$, while the equation of state parameters are $w_{\rm \varphi} = 0$  and $w_{\rm GW, \rm rad} = 1/3 $. The curvature perturbation on uniform total density hypersurfaces is defined as
\begin{equation}
    \zeta \equiv -\Phi - \mathcal{H}\frac{\delta \rho}{\rho^\prime}  \,.
\end{equation}
and it is a weighted sum of the individual perturbations
\begin{equation}
    \zeta = \sum_i \frac{\rho_i^\prime}{\rho^\prime}  \zeta_i\,.
\end{equation}
The curvature perturbation on uniform $i$-component density hypersurfaces is
\begin{equation}
    \zeta_i\equiv -\Phi - \mathcal{H}\frac{\delta \rho_i}{\rho_i^\prime}  \,.
    \label{def:zeta_i}
\end{equation}
We can rewrite Eq. (\ref{con. Eq.}) in terms of the curvature perturbation as
\begin{equation}
    \zeta^\prime_i = \frac{3\mathcal{H}^2\delta P_{\rm{nad},i}}{\rho^\prime_i} - \frac{\mathcal{H}\delta Q_{\rm{nad},i}}{\rho^\prime_i} \,,
\end{equation}
where the first term $\delta P_{\rm{nad},i}$ is the intrinsic non-adiabatic pressure perturbation of each component and it is equal to zero assuming that all fluids have a definite equation of state $P_{i} = P_{i}(\rho_i)$. The second term denotes the non-adiabatic energy transfer for each component. In the case of single inflaton, non-adiabatic energy transfer is absent and the perturbations are adiabatic. Non-adiabatic perturbation cannot arise
on large scales if the initial perturbations were purely adiabatic~\cite{Wands:2000dp}. However, if inflation ends with a period of preheating, the adiabatic condition can be violated due to the strong
parametric resonance. It has been shown in~\cite{Liddle:1999hq} that in this scenario the dominant effect is a second-order term in perturbations of the preheating field and is negligible on large scales. Therefore, the individual curvature perturbations are equal to the total one and remain constant on large scales. Consequently, the perturbations of the components with the same equation of state are equal,
\begin{equation}
    \zeta_{\rm rad}=\zeta_{\rm GW}  \rightarrow \delta_{\rm rad} = \delta_{\rm GW} \, .
\end{equation}
In this case, it is possible to use the $(0,0)$ Einstein equation to determine the initial energy density perturbation. The $(0,0)$ Einstein equation is 
\begin{equation}
   \frac{2}{\mathcal{H}}\left(\Phi^\prime+\mathcal{H}\Psi\right) =  -\frac{\sum_i\bar{\rho}_i\delta_i}{\sum_j\bar{\rho}_j}\, ,
\end{equation}
which gives, when $\delta_i=\delta_j$ for each $i$, $j$,
\begin{equation}
    \delta_i = -2\Psi-2\frac{\Phi^\prime}{\mathcal{H}}.
\end{equation}
The initial perturbation of the distribution function for adiabatic initial conditions of the CGWB has been computed in~\cite{Ricciardone:2021kel,Schulze:2023ich} and it is given by
\begin{equation}
    \Gamma(\eta_{\rm in},\Vec{k},q) = -\frac{2}{4-n_{\rm gwb}(q)}\Psi(\eta_{\rm in},\Vec{k}) \, ,
\end{equation}
where we have neglected the time derivative of $\Phi$, which is subdominant for super-horizon scales. This result is equivalent to the initial condition on the CMB temperature fluctuation $\delta T/T$ when $n_{\rm gwb}=0$. This difference comes from the fact that the spectrum of gravitons is non-thermal, therefore the initial conditions on perturbation of the distribution function keeps track of the scaling in frequency of the monopole signal~\cite{Schulze:2023ich}. In the following section we will show that when the CGWB is not a decay product of the inflaton, there is no a priori reason to set the adiabatic initial conditions, $\zeta_{\rm rad} = \zeta_{\rm GW}$, a different strategy has to be used to compute the initial perturbation of the distribution function.

\section{Non-adiabaticity of gravitons during inflation}
\label{Non-adiabaticity of gravitons during inflation}
We want to compute the initial overdensity of the CGWB in the case of a CGWB produced by quantum fluctuations of the metric during inflation. The initial condition for the CGWB is computed at the time when we start solving the Boltzmann equation, which corresponds to the time $\eta_{\rm in}$ evaluated long after the GWs of frequency $q$ re-enter the causal horizon. According to Big-Bang Nucleosynthesis (BBN) and {\it Planck} constraints~\cite{Caprini:2018mtu} on the amount of extra relativistic species in the early Universe, the amount of energy produced as gravitational waves has the upper bound 
\begin{equation}
\label{rho_gw/rho_rad}
   {\frac{\bar{\rho}_{\rm GW}(\eta_{\rm in})}{\bar{\rho}_{\rm rad}(\eta_{\rm in})}  \lesssim  \frac{g_\star (T_{\rm 0})}{g_\star(T_{\rm in})} \left( \frac{g_{\star \, S} (T_{\rm in})}{g_{\star \, S}(T_{BBN})}\right)^{4/3}\frac{7}{8} \Delta N_{eff}}  \ll 1\, ,
\end{equation}
which implies that the contribution to the energy budget given by primordial GWs is always subdominant.
Under the assumption that only standard radiation and the CGWB are present at $\eta_{\rm in}$, the $(0, 0)$ Einstein's equation, which determines a condition for the energy density of particle species in the Universe, and the scalar traceless part of Einstein's equations, which gives the relation between the scalar perturbations of the metric in terms of the anisotropic stress, are given by
\begin{equation}
    \begin{split}
    \frac{2}{\mathcal{H}}\left(\Phi^\prime+\mathcal{H}\Psi\right) = & -\frac{\bar{\rho}_{\rm rad}\delta_{\rm rad}+\bar{\rho}_{\rm GW}\delta_{\rm GW}}{\bar{\rho}_{\rm rad}+\bar{\rho}_{\rm GW}}\, , \\
    k^2(\Psi-\Phi) =& -12 \mathcal{H}^2\frac{\bar{\rho}_{\rm rad}\pi_{\rm rad}+\bar{\rho}_{\rm GW}\pi_{\rm GW}}{\bar{\rho}_{\rm rad}+\bar{\rho}_{\rm GW}} \, ,
    \end{split}
\end{equation}
with $\pi_{\rm rad}$ and $\pi_{\rm GW}$ the scalar part of the anisotropic stress of radiation and CGWB respectively. Since the energy density of gravitational waves is subdominant w.r.t. standard radiation the Einstein's equations are not sensitive to the presence of the CGWB, therefore no information about $\delta_{\rm GW}$ can be extracted by using the Einstein equations. Additionally, the perturbation of gravitons cannot be directly related to those of photons, because the latter is produced by the decay of the inflaton, while the former by the small-scale perturbations of the metric. Therefore, the only way to get access to the perturbation of gravitons at early times is by directly computing the energy-momentum tensor of GWs in a perturbed FLRW metric~\cite{ValbusaDallArmi:2023nqn}
, exploiting the microscopic description of the field to reconstruct its macroscopic properties.

\section{Inflationary initial conditions for the CGWB}
\label{Inflationary initial conditions for the CGWB}

\subsection{The energy-momentum tensor of the gravitational field}

When the wavelength of GWs is comparable to the scales over which the background varies, the energy-momentum tensor of GWs is ambiguously defined~\cite{Giovannini:2019ioo}. In this case, one could start from the Einstein-Hilbert action~\cite{Ford:1977dj,Ford:1977in} or by considering terms quadratic in the tensor amplitude in the Einstein tensor~\cite{Landau:1975pou}. In~\cite{Babak:1999dc} the GWs are considered as a tensor field on a curved manifold. In this work, we consider only sub-horizon GWs and thus all the definitions of the energy-momentum tensor converge~\cite{Isaacson:1967sln,Misner:1973prb,Landau:1975pou,Giovannini:2019ioo}. For simplicity, we use the definition of the energy-momentum tensor of the CGWB given in~\cite{Isaacson:1967sln}, which consists of a simple expression in terms of covariant derivatives (w.r.t. the slowly-varying metric) of the GWs, 
\begin{equation}
    T_{\mu\nu}^{\rm GW} = \frac{1}{32\pi G}\left\langle\mathcal{D}_\mu \gamma_{\alpha\beta}^{\rm GW}\mathcal{D}_\nu\gamma^{\rm GW\,\alpha\beta}\right\rangle \, .
    \label{def:T_Isaacson}
\end{equation}
The radiative degrees of freedom of the metric in this prescription are identified by the rapidly-oscillating perturbations of the metric, therefore all the terms proportional to $h_{ij}$. In the Poisson gauge we simply have\footnote{Here round parentheses denote symmetrization over indices.} 
\begin{equation}
    \begin{split}
        \gamma_{ij}^{\rm GW} \equiv & a^2 h_{ij} \, , \\
        \gamma^{{\rm GW}\, ij} \equiv &  g^{ik}g^{jl}\gamma_{kl}^{\rm GW} = \frac{1}{a^2}[(1+4\Phi)h^{ij}-H^{k(i}h^{j)}_k] \, .
    \end{split}
\end{equation}
The average, which appears in Eq.~\eqref{def:T_Isaacson}, corresponds to the Brill-Hartle average and it smooths the perturbations on scales much larger than the wavelengths of the GWs and much smaller w.r.t. the scales over which $\Phi$, $\Psi$ and $H_{ij}$ vary~\cite{Isaacson:1967sln,Misner:1973prb}. Additional details on this averaging scheme are given in Appendix \ref{app:BH_average}. As a consistency check, we also compute the energy-momentum tensor by using the approach of~\cite{Landau:1975pou}, by perturbing the Einstein tensor up to ‘‘hybrid third order'' in the perturbations,  
\begin{equation}
        T^{\rm GW}_{\mu\nu} \left(\eta, \Vec{x}\right)=   -\frac{1}{8\pi G} G^{(3)}_{\mu\nu} \left(\eta, \Vec{x}\right)\, ,  
        \label{def:T_Landau}
\end{equation}
where $G^{(3)}_{\mu\nu}$ is quadratic in $h_{ij}$ and up to linear in the large-scale scalar $\Phi, \Psi$ and tensor $H_{ij}$ perturbations.

\subsection{Energy density}

The (0,0) component of the energy-momentum tensor is related to the average GW energy density and its perturbation by
\begin{equation}
       -  T^{0 \, \rm GW}_{0}=   \overline{\rho}_{ \rm GW} +  \delta \rho_{ \rm GW} \, .   
\end{equation}
The covariant derivatives of the gravitational field have been computed in Appendix~\ref{app:Covariant derivatives of GWs} and the energy-momentum tensor computed by using Eq.~\eqref{def:T_Isaacson} is
\begin{equation}
    T^{0\, \rm GW}_0 = -\frac{1}{32\pi G a^2}\left\langle h_{ij}^\prime h^{ij\, \prime}\left(1-2\Psi+4\Phi\right)-2h^\prime_{ij}h^{j\, \prime}_k H^{ik}\right\rangle \, .
    \label{eq:T_00_GW}
\end{equation}
The average GW energy density and its perturbation are therefore~\cite{ValbusaDallArmi:2023nqn}
\begin{equation}
    \label{eq:rho_GW_tot}
    \begin{split}
        \bar{\rho}_{GW}  =& \frac{1}{32\pi G a^2  }\langle h_{ij}' h^{ij\, \prime}\rangle  \, , \\
        \delta\rho_{GW}  =& \frac{1}{32\pi G a^2  }\left[\left\langle h_{ij}' h^{ij\, \prime}\right\rangle\left(-2\Psi+4\Phi\right)-\left\langle h_{ij}^\prime h^{j\, \prime}_k\right\rangle 2H^{ik}\right]  \, ,
    \end{split}
\end{equation} 
where we have used the fact that the Brill-Hartle average does not modify the large-scale perturbations of the metric.

In Eqs.~\eqref{eq:rho_GW_tot} we have computed the energy density of the GWs in terms of the correlation function of $h_{ij}^\prime h^{ij\, \prime}$. However, since $h_{ij}$ is a Gaussian random variable, an object quadratic in the tensor perturbations of the metric is $\chi^2$ distributed. Therefore, $\rho_{\rm GW}$ contains corrections proportional to the four-point function of $h_{ij}$. Nevertheless, these contributions consist in small-scale corrections (they vary just on scales $q$) to the monopole and they do not affect the initial conditions on the anisotropies and can therefore be disregarded.

As a consistency check, we derive the energy-momentum tensor of the gravitational field starting from the Einstein tensor at third order, Eq.~\eqref{def:T_Landau}. To do this we use xPand~\cite{Pitrou:2013hga} to compute the Einstein tensor up to ‘‘hybrid third order'' in perturbations, keeping the terms quadratic in $h_{ij}$ and of order zero and linear in large-scale scalar perturbations $\Phi$, $\Psi$ and $H_{ij}$. The complete expression for the Einstein tensor at third order is rather cumbersome, but the shortwave approximation allows to simplify a large number of terms in the expression. We neglect indeed the friction terms proportional to $\mathcal{H}$ because they are subdominant wr.t. terms proportional to the derivatives of $h_{ij}$, since $q\eta\gg 1$,
\begin{equation}
    \begin{cases}
    \mathcal{H}h_{ij}\approx \frac{1}{\eta} h_{ij} \\
    \partial_\mu h_{ij}  \approx q h_{ij}
    \end{cases}\rightarrow \mathcal{H}h_{ij} \approx \frac{1}{q\eta} \partial_\mu h_{ij} \ll \partial_\mu h_{ij} \, .
\end{equation} 
Spatial and temporal derivatives of the large-scale scalar and tensor perturbations are also negligible at the time at which we fix the initial conditions. After these simplifications, the Einstein tensor at third order is equal to
\begin{equation}
 \begin{split}
     a^2 G^{0\, (3)}_{0} =  \frac{1}{4}\Bigl(&-\Psi h^{ij \,\prime} h_{ij}^\prime+2\Phi h^{ij \,\prime} h_{ij}^\prime-12 \Phi h^{ij} \nabla^2 h_{ij}+6\Phi \partial^k h^{ij}\partial_j h_{ik}-9\Phi \partial^k h^{ij}\partial_k h_{ij}\\
     &- H^{ij}h_i^{k\,^\prime} h_{jk}^\prime-4H^{ij}h^{kl}\partial_j\partial_l h_{ik}+\frac{3}{2}H^{ij}\partial_i h^{kl}\partial_j h_{kl}+2 h^{ij} H^{kl}\partial_k\partial_l h_{ij}\\
     &+4h_i^jH^{ik}\nabla^2 h_{jk}-2H^{ij}\partial_j h_{kl}\partial^l h_i^k- H^{ij}\partial_k h_{jl}\partial^l h_i^k+3 H^{ij}\partial_k h_{jl}\partial^k h_i^l\Bigl)\,. 
 \end{split}   
\end{equation}
As discussed in Appendix~\ref{app:BH_average}, the energy-momentum tensor is defined up to total covariant derivatives quadratic in the radiative degrees of freedom of the metric, see Eq.~\eqref{eq:negligible_total_derivative}. The Einstein tensor can therefore be written as 
\begin{equation}
 \begin{split}
     a^2 G^{0\, (3)}_{0} =  \frac{1}{4}\Bigl(&-2\Psi h^{ij \,\prime} h_{ij}^\prime+4\Phi h^{ij \,\prime} h_{ij}^\prime-2 H^{ij}h_i^{k\,^\prime} h_{jk}^\prime-12\Phi S^{(0)}+6\Phi\mathcal{M}^{(0)}+\frac{1}{2}\mathcal{G}^{(1)} \\&-4H^{ij}\mathcal{A}^{(0)}_{ij}+4H^{ij}\mathcal{P}^{(0)}_{ij}+{+2H^{ij}\mathcal{C}_{ij}^{(0)}}-2H^{ij}\mathcal{D}_{ij}^{(0)}- H^{ij}\mathcal{F}_{ij}^{(0)}\Bigl)\, ,
 \end{split}   
\end{equation}
where the total covariant derivatives have been computed in Eqs.~\eqref{eq:cov_derivatives_scalars},~\eqref{eq:four_divergence}~\eqref{eq:cov_derivatives_tensors}. After neglecting the total covariant derivatives, whose Brill-Hartle average is subdominant, we get an expression consistent with~\eqref{eq:T_00_GW}. In Appendix~\ref{App:The Einstein tensor at second order} we have also shown that from Eq.~\eqref{def:T_Landau} it is possible to obtain the background energy density $\bar{\rho}_{\rm GW}$. 

Analogously, we compute the $(i,j)$ component of the energy-momentum tensor,
\begin{equation}
    T^{i\, \rm GW}_j = \frac{1}{32\pi G a^2}\left[(1+6\Phi)\left\langle \partial^i h_{kl}\partial_j h^{kl}\right\rangle-H^{m(l}\left\langle\partial^i h_{kl}\partial_j h^{k)}_m\right\rangle-H^{im}\left\langle\partial_m h_{kl}\partial_j h^{kl}\right\rangle\right] \, .
\end{equation}
To extract the pressure from the energy-momentum tensor we use just 
\begin{equation}
	\begin{split}
	P_{\rm GW} = & \frac{1}{3}\delta^j_i \, T^{i\, \rm GW}_j =  \frac{1}{32\pi G a^2}\frac{1}{3}\left\langle \partial_r h_{ij} \partial^k h^{ij}\left[(1+6\Phi)\delta^r_k-{H^r_k}\right]-H^{r(l}\partial^s h^{m)}_r\partial_s h_{lm}\right\rangle \, , 
	\end{split}
\end{equation}
where we have exploited the fact that at first order in the large-scale perturbations $\pi_{ij}\delta^{ij}=0$. Up to the four-divergence, Eq.~\eqref{eq:four_divergence}, which is negligible by taking the average, the pressure is 
\begin{equation}
    P_{\rm GW} = \frac{1}{3}\rho_{\rm GW} \, ,
\end{equation} 
which confirms that the CGWB behaves as a radiation fluid also at first order in the perturbations. The $(0,i)$ component of the energy-momentum tensor is just
\begin{equation}
    T^{0\, \rm GW}_i = -\frac{1}{32\pi G a^2}\left[(1-2\Psi+4\Phi)\left\langle \partial_i h_{kl}h^{kl\, \prime}\right\rangle-H^{r(k}\left\langle \partial_i h_{kl}h^{l)\, \prime}_r\right\rangle\right] \, .
\end{equation}
The average velocity of the CGWB is given by
\begin{equation}
	\begin{split}
    v_{i\, \rm GW} =& \frac{1}{\bar{\rho}_{\rm GW}+\bar{P}_{\rm GW}}T^{0\, \rm GW}_i = \frac{3}{4\left\langle h^{lm\, \prime}h_{lm}^\prime\right\rangle}\left[ (1-2\Psi + 4\Phi)\left\langle h^{lm\, \prime}\partial_i h_{lm}\right\rangle-H^{r(l}\left\langle h_{lm}^\prime \partial_i h_{r}^{m)}\right\rangle \rangle\right] \, .
	\end{split}
\end{equation}
Under the assumption that the quantum fluctuations of the tensor perturbations of the metric during inflation preserve statistical isotropy, the average velocity of the CGWB vanishes, since it is proportional to vectors like $ \langle h^{ij\, \prime}\partial_k h_{ij}\rangle$. This expectation value is equal to zero because there is no preferred direction for the average of the single patch of GWs. The anisotropic stress of the CGWB is defined as   
\begin{equation}
   \pi^{i\, \rm GW}_{j} = T^{i\, \rm GW}_j - P^{ \rm GW}\delta_{ij} \, ,
\end{equation}
and thus we find
\begin{equation}
	\begin{split}
	\pi^{i}_{j\, \rm GW} = & \frac{1}{32\pi G a^2}\biggl\langle (1+6\Phi)\left(\partial^i h^{lm}\partial_j h_{lm}-\frac{1}{3}\delta^i_j \partial^k h^{lm}\partial_k h_{lm} \right)\\
     &\hspace{4.5em}-\left(H^{ik}\partial_k h^{lm}\partial_j h_{lm}-\frac{1}{3}\delta^i_j H^{rk}\partial_k h^{lm}\partial_r h_{lm}\right) \\
	&\hspace{4.5em}-\left(H^{rl}\partial^i h^m_r\partial_j h_{lm} - \frac{1}{3}\delta^i_j H^{rl}\partial^sh^m_r\partial_s h_{lm}\right)\biggl \rangle \,.
	\end{split}
\end{equation}
In the next section, we will compute the two-point correlation function of $h_{ij}$ in order to check that it does not contain perturbations of scales $k$ and that all the contributions to the initial conditions on the anisotropies of the CGWB come from $\Phi$, $\Psi$ and $h_{ij}$ in Eq.~\eqref{eq:rho_GW_tot}.

\subsection{Perturbed equation of motion for GWs}

In the previous section we have shown that the energy-momentum tensor of the CGWB contains, according to the definition given by Eq.~\eqref{def:T_Isaacson}, inhomogeneities due to the large-scale perturbations of the metric. These terms come from the definition of covariant and contravariant derivatives and by the fact that the contravariant GWs are defined from the covariant GWs w.r.t. the slowly-varying metric. However, the gravitational radiation $h_{ij}$ could be intrinsically inhomogeneous because of the propagation of the tensor perturbations through the large-scale perturbations of the metric. More explicitly, the equation of motion of GWs is computed by requiring that the transverse-traceless part of the Einstein tensor vanishes,
\begin{equation}
    G^{i\,  (\rm TT)}_j = 0 \, ,
\end{equation}
where we have neglected any source of anisotropic stress. At zero order in the large-scale perturbations the equation of motion is 
\begin{equation}
    h_{ij}^{\prime\prime}+2\mathcal{H}h_{ij}^\prime-\nabla^2 h_{ij} = 0 \, .
    \label{eq:eom_GWs_unperturbed}
\end{equation}
On the other hand, when the presence of $\Phi$, $\Psi$ and $H_{ij}$ is taken into account, the full expression of the equation of motion is 
\begin{equation}
    \begin{split}
    & \left(1-2\Psi+2\Phi\right)\Bigl[h_{ij}^{\prime\prime}+h_{ij}^\prime\left(2\mathcal{H}-\Psi^\prime+\Phi^\prime\right)+h_{ij}\left(4\mathcal{H}\Phi^\prime+2\Phi^{\prime\prime}\right)-\nabla^2 h_{ij}\left(1+2\Psi+2\Phi\right)\\
    &\hspace{8.35em}+2H^{kl}\partial_k\partial_l h_{ij}\Bigl] = 0  \, .
    \label{eq:perturbed_eom_GWs}
    \end{split}
\end{equation}
To compute it, we have to evaluate the transverse-traceless part of the Einstein tensor at second order, neglecting terms quadratic in the perturbations, because they would act as a source of scalar-induced GWs~\cite{1971PThPh..45.1747T,Matarrese:1992rp,Matarrese:1993zf,Matarrese:1997ay} not considered in this work. If we neglecr $H_{ij}$ and assume that $\Phi$ and $\Psi$ are constant in time and uniform in space, it is possible to treat them as constants and simplify the equation of motion,
\begin{equation}
    h_{ij}^{\prime\prime}+2\mathcal{H}h_{ij}^\prime-\nabla^2 h_{ij}\left(1+2\Psi+2\Phi\right) = 0 \, .
\end{equation}
It is possible to decompose the GWs as 
\begin{equation}
    h_{ij}(\eta,\Vec{x}) = \int \frac{d^3q}{(2\pi)^3}\sum_{\lambda = \pm 2} e^{i\Vec{q}\cdot \Vec{x}}\, e_{ij}^\lambda(\hat{q}) h_\lambda(\Vec{q})T_h(\eta,q) \, ,
\end{equation}
where $T_h(\eta,q)$ is the transfer function of the tensor perturbations of momentum $q$ at the time $\eta$, $h_\lambda(\Vec{q})$ is the primordial tensor perturbation and $\lambda = \pm 2$ accounts for the two polarisation degrees of freedom. During radiation-dominated epoch, $\mathcal{H}=1/\eta$, and the equation of motion of the transfer function reads
\begin{equation}
    T_h^{\prime\prime}(\eta, \Vec{x}, q)+\frac{2}{\eta}T_h^\prime(\eta, \Vec{x}, q)+q^2\left(1+2\Phi+2\Psi\right)T_h(\eta, \Vec{x}, q) = 0 \, .
\end{equation}
The general solution of the equation is given by
\begin{equation}
    T_h(\eta, \Vec{x}, q) = j_0\left(Q(\eta,\Vec{x}, q)\, \eta\right) \, ,
\end{equation}
where $ j_0$ is the spherical Bessel function of zeroth order and  
\begin{equation}
Q(\eta,\Vec{x}, q)\equiv \left[1+\Psi(\eta,\vec{x})+\Phi(\eta,\vec{x}) \right]q\, .
\end{equation}
The energy density of GWs is proportional to the temporal derivative of the small-scale mode, then
\begin{equation}
    T_h^\prime(\eta, \Vec{x}, q)= -Q(\eta, \Vec{x}, q) j_1\left(Q( \Vec{x}, q) \eta\right) \, .
\end{equation}
If we set the initial condition when the modes are inside the horizon, $q\eta \approx 150$, we can write explicitly the spherical Bessel function of order one  as
\begin{equation}
    T_h^\prime(\eta, \Vec{x}, q) = -\frac{1}{\eta}\cos\left(Q(\eta, \Vec{x}, q)\eta\right)\approx -\frac{1}{\eta}\cos(q\eta)+\sin(q\eta)\left(\Psi(\eta,\vec{x})+\Phi(\eta,\vec{x})\right)\, .
\end{equation}
The second term represent a large-scale perturbation of the GWs due to the propagation across underdense and overdense regions. Since we are interested in computing the initial condition of the CGWB, the quantity which enters in Eq.~\eqref{eq:rho_GW_tot} is the average of the square of the transfer function,
\begin{equation}
    \langle \left[T_h^\prime(\eta, \Vec{x}, q)\right]^2\rangle \approx \frac{1}{\eta^2}\langle \cos^2(q\eta)\rangle+2\left(\Psi(\eta,\vec{x})+\Phi(\eta,\vec{x})\right)\langle \sin(2q\eta)\rangle = \frac{1}{2\eta^2} \, ,
\end{equation}
where the average has been taken w.r.t. $q$. Since the average of the sine is zero, the perturbations in $h_{ij}$ due to the equation of motion are negligible and for the purposes of this paper, we can consider $h_{ij}$ as a tensor linear in the small-scale perturbations and of order zero in the large-scale perturbations.

\subsection{Initial condition term for the CGWB}
The energy density, pressure, velocity and anisotropic stress of the CGWB are evaluated by integrating over all the frequency spectrum of the CGWB because the GWs considered $h_{ij} (\eta, \Vec{k})$ are the superposition of many small-scale tensor perturbations of frequencies $q$. The 2-point function for tensor perturbations at a given time $\eta$ is 
\begin{equation}
    \left\langle h_{\lambda}(\eta,\Vec{q})h_{\lambda^\prime}(\eta,\Vec{q}^\prime)\right\rangle = \frac{8\pi^5}{2q^3}P_T(q)\, \delta_{\lambda\lambda^\prime}\, \delta(\Vec{q}-\Vec{q}^\prime) \, \left[T_h(\eta,q)\right]^2 \, .
\end{equation}
The energy density of the CGWB then is 
\begin{equation}
    \begin{split}
    &\rho_{\rm CGWB}(\eta_{\rm in},\vec{x}) = \frac{1}{32\pi G a^2 \left(\eta\right)} \int \frac{d^3q}{(2\pi)^3} \frac{2\pi^2}{q^3}P_T(q)\left\langle \left[T_h^\prime(\eta_{\rm in},q)\right]^2\right\rangle \\
    &\left[1-2\Psi(\eta_{\rm in},\Vec{x})+4\Phi(\eta_{\rm in},\Vec{x})-2 H^{ij}(\eta_{\rm in},\Vec{x})\sum_{\lambda = \pm 2} e^\lambda_{ik}(\hat{q})e_j^{k\, \lambda}(\hat{q})\right] \, .
    \end{split}
\end{equation}
It is possible to show that
\begin{equation}
    \sum_\lambda e^\lambda_{ik}(\hat{q})e^{k\, \lambda}_j(\hat{q}) = \delta_{ij}-\hat{q}_i\hat{q}_j \, , 
    \label{eq:pol_tensor_rule}
\end{equation} 
then
\begin{equation}
    \begin{split}
    \rho_{\rm CGWB}(\eta_{\rm in},\vec{x}) = & \frac{1}{32\pi G a^2\left(\eta_{\rm in}\right)} \int \frac{d^3q}{(2\pi)^3} \frac{2\pi^2}{q^3}P_T(q)\left\langle \left[T_h^\prime(\eta_{\rm in},q)\right]^2\right\rangle \\&\left[1-2\Psi(\eta_{\rm in},\Vec{x})+4\Phi(\eta_{\rm in},\Vec{x}) + 2 H^{ij}(\eta_{\rm in},\Vec{x})\hat{q}_i\hat{q}_j\right] \, .
    \end{split}
\end{equation}
It is immediate to connect the isotropic component of the graviton distribution function to the primordial tensor power spectrum
\begin{equation}
    \bar{f}(q) = \frac{\pi}{16 G a^2(\eta_{\rm in}) q^3}P_T(q)\left\langle\left[T_h^\prime(\eta_{\rm in},q)\right]^2\right\rangle \, .
\end{equation}
and the anisotropic contribution to the energy density of GWs is
\begin{equation}
\begin{split}
        \delta_{\rm GW}(\eta_{\rm in},\Vec{k}, \hat{n}, q)&=-2\Psi(\eta_{\rm in},\Vec{k})+4\Phi(\eta_{\rm in},\Vec{k}) 
      + 2 \sum_{\lambda = \pm 2} H_\lambda(\eta_{\rm in},\Vec{k}) e^{ij}(\hat{k})\hat{q}_i\hat{q}_j\, .
\end{split}
\end{equation}
Assuming that the primordial tensor modes are unpolarized we can use Eq. \eqref{eq:pol_tensor_rule}, then the perturbation of the energy density is
\begin{equation}
\begin{split}
      \delta_{\rm GW}(\eta_{\rm in},\Vec{k}, \hat{n}, q) &= -2\Psi(\eta_{\rm in},\Vec{k})+4\Phi(\eta_{\rm in},\Vec{k})
     +2 H(\eta_{\rm in},\Vec{k})\left(1-\mu^2\right)\, ,   
\end{split} 
\end{equation}
which is equivalent to 
\begin{equation}
    \Gamma(\eta_{\rm in},\Vec{k}, \hat{n}, q) = \frac{1}{4-n_{\rm gwb}(q)}\left[-2\Psi(\eta_{\rm in},\Vec{k})+4\Phi(\eta_{\rm in},\Vec{k})
     +2 H(\eta_{\rm in},\Vec{k})\left(1-\mu^2\right)\right] \, .
     \label{eq:iic_Gamma}
\end{equation}
We can decompose the initial anisotropy of the CGWB into multipoles
   \begin{equation}
   \begin{split}
     \delta_{\rm GW}(\eta_{\rm in},\Vec{k}, \hat{n}, q) = \sum_{\ell} (-i)^{\ell} (2\ell + 1) \delta_{\rm{GW},\ell}(\eta_{\rm{in}},\Vec{k},q)P_{\ell}\left(\mu\right) \, ,   
       \end{split}
   \end{equation}
where $P_{\ell}\left(\mu\right)$ are the Legendre polynomials. The dipole will be zero due to the fact that the average velocity of the CGWB is zero. The quadrupole is sourced by large scale tensor perturbations and therefore it is non-zero but remains comparatively small w.r.t. the monopole. We find that the initial anisotropy of the CGWB is the sum of a monopole and a quadrupole,
\begin{equation}
    \begin{split}
    \delta_{\rm{GW},0}(\eta_{\rm in},\Vec{k}, \hat{n}, q)  &= -2\Psi(\eta_{\rm{in}},\Vec{k})+4\Phi(\eta_{\rm{in}},\Vec{k})  +\frac{2}{3}H(\eta_{\rm{in}},\Vec{k}) \, , \\
    \delta_{\rm{GW},1}(\eta_{\rm in},\Vec{k}, \hat{n}, q)  &=  0 \, , \\
   \delta_{\rm{GW},2}(\eta_{\rm in},\Vec{k}, \hat{n}, q)  &=  -\frac{2}{3}H(\eta_{\rm{in}},\Vec{k}) \, .
    \end{split}
\end{equation}

\section{The angular power spectrum of the CGWB}
\label{sec:The angular power spectrum of the CGWB}
For simplicity, we consider a CGWB produced by quantum fluctuations of the metric in the case of single-field inflation or any inflationary mechanism that gives rise to adiabatic initial conditions for radiation and matter perturbations. The angular power-spectrum of the CGWB defined in Eq. (\ref{power-spectrum}) includes the initial condition term and the propagation effects evaluated in~\eqref{eq:Gamma0_solution}. The redshift at the start of the free-streaming of gravitons gives rise to the Sachs-Wolfe (SW) effect, while the propagation through the large-scale scalar and tensor perturbations to the Integrated Sachs-Wolfe (ISW)~\cite{Contaldi:2016koz,Bartolo:2019oiq,Bartolo:2019yeu}. Following~\cite{Schulze:2023ich} we write the angular power spectrum in terms of the source functions
  \begin{equation}
  \begin{split}
  \frac{C^{\rm GW}_{\ell}}{4\pi(4-n_{\rm gwb})^2} =  \int \frac{dk}{k} \Bigl[&
   P_{\mathcal{R}}(k) \left(\Delta^{I-S}_\ell+\Delta_\ell^{\rm SW}+\Delta_\ell^{\rm ISW}\right)^2+ P_{T}(k) \left(\Delta^{I-T}_\ell+\Delta_\ell^{{\rm ISW}-T}\right)^2\Bigl] \, ,
   \end{split}
   \end{equation}
where $P_\mathcal{R}$ and $P_T$ are the power spectra of scalar and tensor perturbations. 

Neglecting the tensor perturbations, we can see that the contribution to the CGWB anisotropy from these inflationary initial conditions (IIC) enhances the angular power-spectrum compared to the adiabatic case,
\begin{equation}
\begin{split}
      \frac{C_\ell^{\rm IIC+SW}}{C_\ell^{\rm AD+SW}} &\approx  \left(\frac{4T_\Phi+[2-n_{\rm gwb}(q)]T_\Psi}{[2-n_{\rm gwb}(q)]T_\Psi}\right)^2\, .
\end{split}
\end{equation}
When $n_{\rm gwb}=0$, the angular power spectrum is enhanced by a factor of 10, therefore the IIC could enhance the detectability of the anisotropies by at least one order of magnitude. These large initial conditions would allow to overcome the intrinsic variance limit in the detection of the CGWB computed in~\cite{Mentasti:2023icu,Mentasti:2023gmg}.

\begin{figure}[t!]
    \centering
    \includegraphics[width=0.49\textwidth]{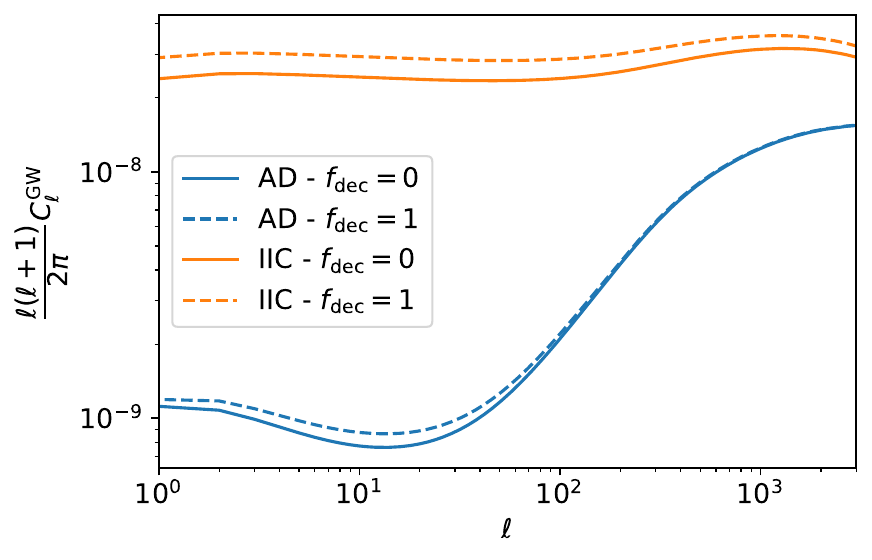}
    \includegraphics[width=0.49\textwidth]{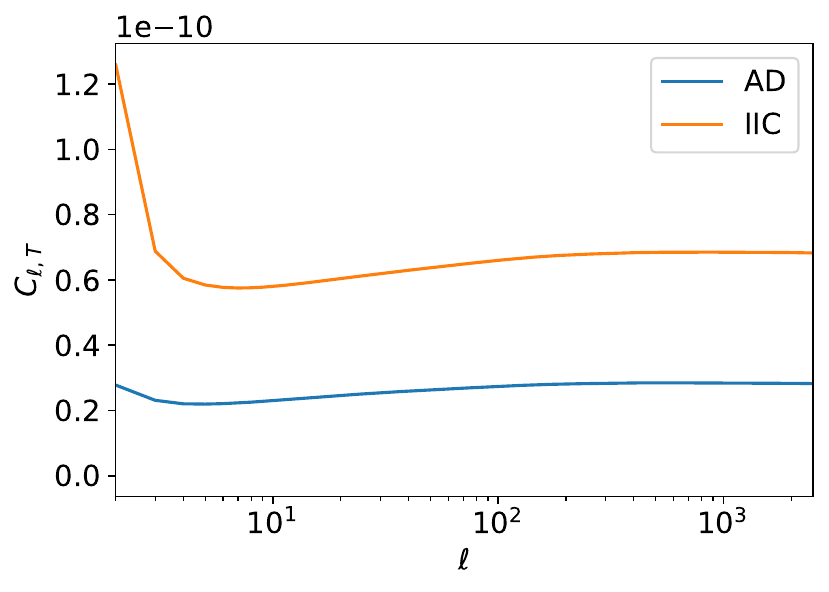}
    \caption{Left: plot of the angular power spectrum of the CGWB for adiabatic initial conditions (AD) and for inflationary initial conditions (IIC) with $n_{\rm gwb} = 0.35$. Right: plot of the tensor contributions to the angular power spectrum for AD and IIC.}
    \label{fig:AD_IIC}
\end{figure}

Since the total angular power-spectrum is the sum of the initial condition term, the SW, the ISW effect and the cross-correlation between these terms, differences in the initial conditions could change the way in which these three effects combine together, modifying also the scaling of the angular power spectrum. For instance, the adiabatic initial conditions have the same sign of the early ISW and thus it sums coherently at large $\ell$, generating a bump in the spectrum, while the IIC have the opposite sign and the spectrum remains almost constant when $\ell$ increases. Furthermore, the IIC could change also the features of the angular power spectrum at high multipoles, since the initial condition adds differently to the SW and the ISW effect w.r.t. the adiabatic case. In the left panel of Fig. \ref{fig:AD_IIC} we plot the angular power-spectrum of the CGWB for IIC and for adiabatic initial conditions with $n_{\rm gwb} = 0.35$. The spectrum has been computed with a modified version of the \texttt{GW$\_$CLASS} code which includes the non-adiabatic initial conditions computed in Eq.~\eqref{eq:iic_Gamma}. The flattening of the angular power spectrum in the case of large initial conditions due to scalar isocurvature perturbations has also been discussed in~\cite{Malhotra:2022ply}.

It is known that the angular power-spectrum of the CGWB is sensitive to the relativistic and decoupled degrees of freedom~\cite{ValbusaDallArmi:2020ifo, Schulze:2023ich}. The impact of these parameters depends on the choice of the initial conditions. The transfer functions of the scalar metric fluctuations depend on the fractional energy density of relativistic and decoupled particles species through 
\begin{equation}
    \begin{split}
    T_\Psi(\eta_{\rm in},k) =& -\frac{2}{3}\left(1+\frac{4}{15}f_{\rm dec}(\eta_{\rm in})\right)^{-1} \, , \\
    T_\Phi(\eta_{\rm in},k) = & \left[1+\frac{2}{5}f_{\rm dec}(\eta_{\rm in},k)\right]T_\Psi(\eta_{\rm in},k) \, .
    \end{split}
\end{equation}
The dependence on the number of relativistic and decoupled particles species enters in the source function for the initial condition, for the SW and for the primordial ISW, integrated between $\eta_{\rm in}$ and  $\eta_{\rm min}$ 
\begin{equation}
    \begin{split}
    \Delta^{\rm SW}_\ell(\eta_0,k,q,\eta_{\rm in}) = & \,  T_\Psi(\eta_{\rm in},k)j_\ell[k(\eta_0-\eta_{\rm in})] \, , \\
    \Delta^{\rm ISW-prim}_\ell(\eta_0,k,q,\eta_{\rm in}) = &   j_{\ell}(k(\eta_0 - \eta_{\rm in}))\left[T_{\Psi}(\eta_{\rm min},k)+ T_{\Phi}(\eta_{\rm min},k)-T_{\Psi}(\eta_{\rm in},k)- T_{\Phi}(\eta_{\rm in},k)\right] \\
& \approx  - j_{\ell}(k(\eta_0 - \eta_{\rm in}))\left[2+\frac{2}{5}f_{\rm dec}(\eta_{in}))\right]T_{\Psi}(\eta_{\rm in}) \, .
    \end{split}
\end{equation}
For adiabatic initial conditions, it is possible to show that, at large angular scales, the angular power spectrum of the CGWB depends on $f_{\rm dec}(\eta_{\rm in})$ through the combination
\begin{equation}
    C_\ell^{\rm AD+SW+ISW-prim}\approx\left(\frac{\frac{6-n_{\rm gwb}}{4-n_{\rm gwb}}+\frac{2}{5}f_{\rm dec}(\eta_{\rm in})}{1+\frac{4}{15}f_{\rm dec}(\eta_{\rm in})}\right)^2 \, .
    \label{eq:Cell_fdec_ad}
\end{equation}
When $n_{\rm gwb}=0$, the angular power spectrum is independent of $f_{\rm dec}(\eta_{\rm in})$. Anyway, in most cases for the CGWB potentially detectable by future interferometers $n_{\rm gwb}$ will be different from zero. For a spectrum with $n_{\rm gwb} = 0.35$, a higher number of relativistic species suppresses the sensitivity of the anisotropies to the relativistic degrees of freedom due to the compensation of the SW effect and early ISW effect with the adiabatic initial conditions. On the other hand, for the IIC we find
\begin{equation}
    C_\ell^{\rm IIC+SW+ISW-prim}\approx\left(\frac{\frac{2-n_{\rm gwb}}{4-n_{\rm gwb}}-\frac{2}{5}f_{\rm dec}(\eta_{\rm in})\frac{n_{\rm gwb}}{4-n_{\rm gwb}}}{1+\frac{4}{15}f_{\rm dec}(\eta_{\rm in})}\right)^2 \, ,
    \label{eq:Cell_fdec_SFI}
\end{equation}
which means that when $n_{\rm gwb}=0$, the spectrum still depends on $f_{\rm dec}(\eta_{\rm min})$. Note also that in the adiabatic case an increase of $f_{\rm dec}(\eta_{\rm in})$ determines a suppression of the angular power spectrum, while in the case of IIC the presence of the relativistic particles enhances the anisotropies. Moreover, in the neighbourhood of $n_{\rm gwb}=0$, the enhancement of the angular power spectrum of the CGWB for change in $f_{\rm dec}(\eta_{\rm in})$ is much more visible than the suppression in the adiabatic case. This occurs because in Eq. (\ref{eq:Cell_fdec_ad}) there is a partial cancellation between the dependence on $f_{\rm dec}(\eta_{\rm in})$ between the numerator and the denominator, while in Eq. (\ref{eq:Cell_fdec_SFI}) this does not occur. In the left panel of Figure~\ref{fig:AD_IIC} we have computed the angular power spectrum for $f_{\rm dec}(\eta_{\rm in})=0$ and $f_{\rm dec}(\eta_{\rm in})=1$, showing that in the case of IIC the spectrum is more sensitive to relativistic species.

\section{Tensor contribution to the angular power spectrum of the CGWB}
\label{Tensor contribution to the angular power spectrum}
In the standard case~\cite{Schulze:2023ich}, the tensor perturbations contribute to the angular power spectrum of the CGWB through the ISW effect, due to the variations of $H_{ij}$ in time,
\begin{equation}
    \Gamma^{\rm{ISW} - T}(\eta_0,k,q) = -\frac{1}{2}\int_{\eta_{\rm in}}^{\eta_0}d\eta e^{ik\mu(\eta-\eta_0)}H_{ij}^\prime \hat{q}^i\hat{q}^j \, .
\end{equation}
In the case of IIC, this contribution has to be summed with the tensor term related to the initial conditions,
\begin{equation}
    \Gamma^{I-T}(\eta_{\rm{in}},\Vec{k},q) = \frac{2}{4-n_{\rm gwb}(q)}e^{ik\mu(\eta_{\rm in}-\eta_0)}H_{ij}\hat{q}^i\hat{q}^j \, .    
\end{equation}
The two terms considered have the same sign, since the time derivative of $H_{ij}$ is negative, which means that the two contributions add coherently, giving rise to a larger anisotropy. The net effect is to increase by an order of magnitude the tensor contribution as for the scalar part. We note however that the primordial tensor power spectrum is suppressed w.r.t. the scalar one by the small value of the tensor-to-scalar ratio constrained by \textit{Planck} and BICEP, $r\lesssim 0.03$~\cite{Planck:2018jri,Tristram:2021tvh,Galloni:2022mok}, therefore we expect that the tensor contribution to the angular power specturm of the CGWB will be subdominant. In the right panel of Figure~\ref{fig:AD_IIC}, we plot the total tensor contribution (the superposition of the initial conditions and the ISW) for adiabatic and inflationary initial conditions.

\section{Cross-correlation between the CGWB and the CMB}
\label{Cross-correlation between the CGWB and the CMB}
At low multipoles, the angular power-spectrum of the CGWB and of the CMB are dominated by the combination of the initial condition and the SW, which is the same for both photons and gravitons, since at large angular scales the two signals experience the redshift from highly-correlated metric perturbations~\cite{Ricciardone:2021kel}. The constraints on the angular power-spectrum of CMB anisotropies show that the initial conditions for photons are adiabatic~\cite{Planck:2018jri}. Therefore, when we consider adiabatic initial conditions of the CGWB, the correlation is very large. On the other hand, the initial conditions in the case of single-field inflation are still correlated with large scale CMB anisotropies at last scattering, because the scalar perturbations that generate them arise from the same seeds, but since the initial conditions combine differently with the other contributions to the angular power spectrum, the correlation decreases by a small fraction. The adiabatic initial term plus the SW effect for the CMB is 
\begin{equation}
    T_\Psi(\eta_*,k)+\Theta_0(\eta_*,k) = \frac{1}{3}T_\Psi(\eta_*,k) = \frac{1}{3}\frac{9}{10}T_\Psi(\eta_{\rm in},k) = \frac{3}{10}T_\Psi(\eta_{\rm in},k) \, .
\end{equation}
For the CGWB the adiabatic initial term plus the SW effect is proportional to $T_\Psi(\eta_{\rm in},k)/2$, which is close to the analogue term for the CMB, while in the case of IIC we have an additional term, which gives a substantial deviation between the adiabatic case because of the way how the adiabatic initial condition term and the SW effect combines with the late ISW effect. This results in a decrease of the correlation of the CMB and of the CGWB defined by 
\begin{equation}
    r_\ell^{\rm CMB\times CGWB} \equiv \frac{C_\ell^{\rm CMB\times CGWB}}{\sqrt{C_\ell^{\rm CMB\times CMB}C_\ell^{\rm CGWB\times CGWB}}} \, .
\end{equation}
For the adiabatic initial conditions we get $r\approx 0.98$, while in the case of IIC we have $r\approx 0.8-0.9$. Thus, the correlation between the CMB and the CGWB  would be a way to test the nature of the initial conditions, as soon as a CGWB is detected. We plot this result in Fig. \ref{fig:r}.

\textbf{\begin{figure}[t!]
    \centering
    \includegraphics[width=0.49\textwidth]{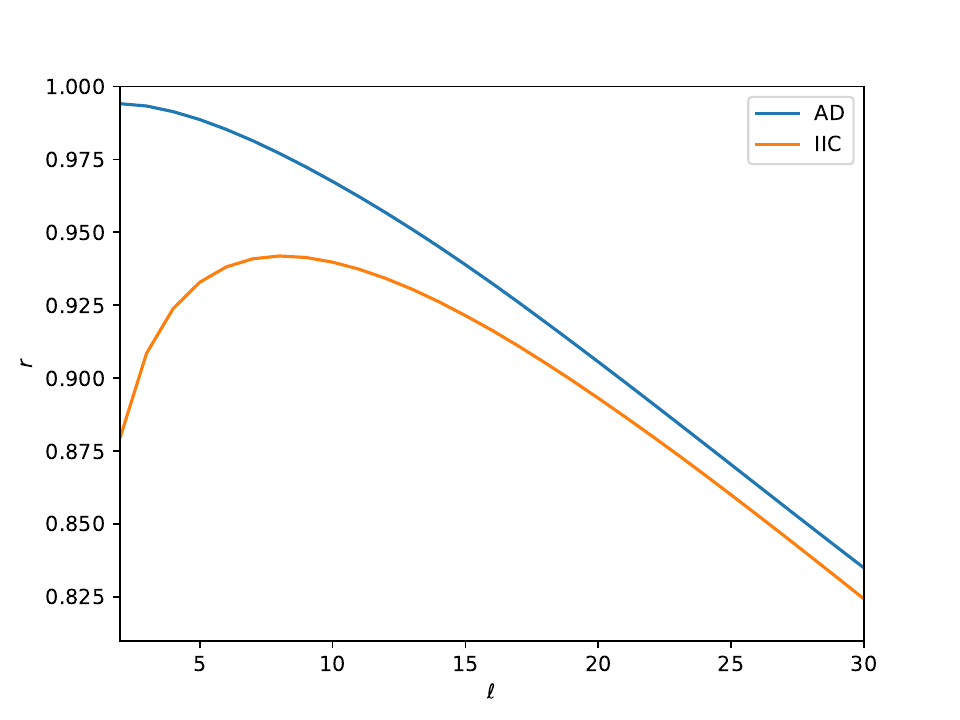}
    \caption{Plot of the correlation between the CMB and the CGWB at different multipoles for adiabatic initial conditions (AD) and for inflationary initial conditions (IIC) with $n_{\rm gwb} = 0.35$.}
    \label{fig:r}
\end{figure}}

\section{An example: the curvaton mechanism}
\label{An example: Curvaton mechanism}

The initial conditions of the CGWB computed in this work apply to any CGWB generated by the quantum fluctuations of the metric independent of the fluctuations of the fields that generate the scalar perturbations of the Universe. An alternative way to say this is that Eq.~\eqref{eq:iic_Gamma} holds even if more than one scalar field is present during inflation. The initial overdensity of gravitons depends indeed just on the large-scale perturbations of the metric at the initial time. Neglecting the large-scale tensor modes and assuming that $f_{\rm dec}(\eta_{\rm in}) = 0$, the energy density perturbation of the CGWB can be written in terms of the total curvature perturbation as 
\begin{equation}
    \frac{\delta\rho_{\rm GW}}{\bar{\rho}_{\rm GW}}= -2\Psi+4\Phi\approx 2\Phi \approx -\frac{4}{3}\zeta\,,
    \label{eq:iic_zeta}
\end{equation}
where the connection between the curvature perturbation and the large-scale perturbations during the radiation dominated epoch is given by $\Psi=\Phi=-2/3\, \zeta$. \\
In this section we show how the IIC apply also in the case of the curvaton mechanism. In the curvaton scenario, the curvature perturbations are not generated by the inflaton, but by the fluctuations of another light scalar field, the curvaton. During inflation, the curvaton plays a negligible role, while in the radiation dominated Universe, before its decay, it gives the dominant contribution to the curvature and to the energy density. We consider the case in which the inflaton and  the curvaton (during the radiation-dominated epoch) decay into standard radiation, while the CGWB is generated by the quantum fluctuations of the metric~\cite{Malhotra:2022ply}. Immediately before the decay of the curvaton we assume that the total curvature perturbations is
\begin{equation}
        \zeta_{\rm in} = f_{\chi, \rm in} \zeta_{\chi,\rm in}+f_{\rm GW,  \rm in}\zeta_{\rm GW,in}+(1-f_{\chi, \rm in}-f_{\rm GW, \rm in})\zeta_{r, \rm in}\, .
    \label{eq:zeta_b}
\end{equation}
where we have defined
\begin{equation}
    f_i \equiv \frac{(1+w_i)\bar{\rho}_i}{\sum_j (1+w_j)\bar{\rho}_j}\, ,
\end{equation}
with $w_j$ the equation of state of the species $j$. Right after the decay of the curvaton into radiation, the curvature perturbation is 
\begin{equation}
    \zeta = f_{\rm GW}\zeta_{{\rm GW}}+\left(1-f_{\rm GW}\right)\zeta_{r} \, .
    \label{eq:zeta_a}
\end{equation}
In the limit of instantaneous decay, the curvaton completely dominates the energy density before it decays. In this case, the total curvature perturbation is approximately constant at the decay epoch and therefore the total curvature perturbation after the decay is proportional to the total curvature perturbation immediately before the decay, $\zeta = \zeta_{\rm in}$\footnote{Before the decay of the curvaton, the curvature perturbation grows in time because $f_{\chi,\rm in}$ increases. This leads to the non-zero non-adiabatic pressure, but this is not important for our purposes.}. We can also use $\zeta_{{\rm GW}} = \zeta_{{\rm GW},\rm in}$ and $f_{\rm GW}= f_{\rm GW, in}$, because the energy density of the CGWB is conserved. In most of the cases, we expect that the contribution to the energy budget of the Universe in form of GWs is subdominant, thence $f_{\rm GW}\ll 1$. Moreover, we assume that the curvaton dominates the curvature perturbation before the decay, $\zeta_{r, \rm in}\ll \zeta_{\chi,\rm in}$. By neglecting the contributions proportional to $f_{\rm GW}$ and $f_{\rm GW, in}$ in Eqs.~\eqref{eq:zeta_b},~\eqref{eq:zeta_a} it is possible to relate the initial conditions on the curvature perturbation of standard radiation and the curvaton,
\begin{equation}
    \zeta_{\rm in} \simeq f_{\chi, \rm in} \zeta_{\chi,\rm in} \, \rightarrow  \zeta_{r} \simeq  \zeta \simeq \zeta_{\rm in}  \simeq f_{\chi, \rm in} \zeta_{\chi,\rm in} \, .
\end{equation}
The quantity $f_{\chi,\rm in}\zeta_{\rm \chi, in}$ is related to the scalar power spectrum and it is constrained by \textit{Planck}~\cite{Planck:2018jri}. The perturbation of the temperature of radiation is then 
\begin{equation}
    \frac{\delta T_r}{T_r} = \frac{1}{4}\frac{\delta\rho_r}{\bar{\rho}_r} = \Phi+\zeta_r \simeq -\frac{2}{3}f_{\chi,\rm in}\zeta_{\chi,\rm in}+f_{\chi,\rm in}\zeta_{\chi,\rm in} \simeq \frac{1}{3}f_{\chi,\rm in}\zeta_{\chi,\rm in} \, .
\end{equation}
As we have shown in this work, the adiabatic assumption is not valid when the gravitons and radiation are originated by different fields, therefore $\zeta_{{\rm GW},\rm in} \neq \zeta_{r,\rm in}$. A self-consistent computation, which follows from Eq.~\eqref{eq:iic_zeta}, gives, for $n_{\rm gwb}=0$,
\begin{equation}
    \frac{\delta\rho_{\rm GW}}{\bar{\rho}_{\rm GW}} \simeq -\frac{4}{3}f_{\chi,\rm in} \zeta_{\chi,\rm in}\rightarrow \Gamma \simeq -\frac{1}{3}f_{\chi,\rm in}\zeta_{\chi,\rm in} \, .
    \label{eq:perturbation_energy_density_curvaton_correct}
\end{equation}
In this way it possible to extract from the energy density of the CGWB also the relative curvature perturbation of gravitons,
\begin{equation}
    \zeta_{{\rm GW}} = \frac{1}{4}\frac{\delta\rho_{\rm GW}}{\bar{\rho}_{\rm GW}}-\Phi \simeq \frac{1}{3}f_{\chi, \rm in }\zeta_{\chi,\rm in} \, .
\end{equation}

\section{Conclusions}

The decoupling of gravitational waves at the Planck epoch makes any stochastic background of gravitational waves sensitive to the physics operating at the energy scales of its production. Since future detectors will have a much better angular resolution, anisotropies of the CGWB could be detected and provide an important tool to  disentangle different astrophysical and cosmological contributions to the SGWB. The anisotropies are imprinted at the production of the CGWB and during the propagation of gravitons through the perturbed Universe. The initial anisotropies could be extremely important, because they retain crucial information about the primordial mechanisms that generates the gravitational waves. However, the definition of the initial overdensity of the CGWB can be challenging, because the distribution of gravitons is not thermal. In particular, the adiabatic initial condition, that holds for CMB photons, is not guaranteed a priori for the primordial GWs. In this paper, we have shown that any CGWB generated by the quantum fluctuations of the metric during inflation has an intrinsic non-adiabatic perturbation, even in single-field models of inflation. The violation of adiabaticity is due to the presence of independent tensor perturbations of the metric during inflation, which fluctuate independently of the dominant source of scalar perturbations of the Universe, such as the inflaton. 

In the case of cosmological background generated by the quantum fluctuations of the metric, the initial conditions are different w.r.t. the CMB ones, because the latter is a thermalized species and a decay product of the fields responsible for the curvature of the Universe. Therefore, no information on the initial conditions on the anisotropies of the CGWB can be extracted from the analogy with the CMB. Furthermore, we have shown that the Einstein equations are insensitive to the presence of the CGWB, because the energy density of primordial gravitons is subdominant w.r.t. standard radiation. Thus, we concluded that the only way possible to evaluate the initial energy density perturbation of the CGWB is by computing the energy-momentum tensor of GWs directly in terms of the gravitational field. To do this, we perturb the energy-momentum tensor of the GWs given in terms of covariant derivatives of the radiative degrees of freedom of the metric w.r.t. the large-scale scalar and tensor perturbations of the Universe. As a consistency check, we perturb also the Einstein tensor up to second order in small-scale tensor modes and linear in large-scale metric perturbations. We find that the initial conditions computed with the two approaches are equivalent and that the non-adiabatic contribution to the energy density of a CGWB consists in a monopole term, proportional to the scalar and tensor perturbations, and in a quadrupole term, proportional just to the tensor perturbations. The scalar part of the initial conditions is very large and its combination with the other contributions to the anisotropies lead to an enhancement of the CGWB angular spectrum by about one order of magnitude w.r.t. the adiabatic case. In addition, the non-adiabatic initial condition increases the sensitivity of the anisotropies to the presence of extra relativistic degrees of freedom in the early Universe. On the other hand, the correlation between the CGWB and the CMB decreases due to the fact that these initial conditions combine differently with the other
contributions to the angular power spectrum. 

The computation of the initial conditions for the CGWB done in this work represents a turning point for the study of the anisotropies of cosmological relics produced by non-thermal and non-adiabatic mechanisms. The arguments we gave to justify the non-adiabaticity of primordial gravitons could indeed be applied to other cosmological relics generated by the quantum fluctuations of scalar fields independent of the inflaton or other fields that dominate the curvature perturbation of the Universe at early times. This has been shown, for instance, in the example of a CGWB generated by the quantum fluctuations of the metric in the curvaton scenario. 

Our findings could have then important implications for the detectability of the anisotropies by future interferometers, since the non-adiabatic initial conditions computed in this work augment the amplitude of the angular power spectrum and the sensitivity to some cosmological parameters by at least one order of magnitude. In addition, differences in the initial conditions of different cosmological (or astrophysical) backgrounds could be an important tool to distinguish among different sources of GWs in the future, once these anisotropies will be detected.

\acknowledgments

We thank N. Bartolo, D. Bertacca, I. Caporali, and A. Greco for useful discussions. S.M. acknowledges partial financial support by ASI Grant No. 2016-24-H.0. A.R. and L.V. acknowledge financial support from the Supporting TAlent in ReSearch@University of Padova (STARS@UNIPD) for the project “Constraining Cosmology and Astrophysics with Gravitational Waves, Cosmic Microwave Background and Large-Scale Structure cross-correlations’'.

\appendix

\section{Brill-Hartle average}

\setcounter{equation}{0}
\renewcommand{\theequation}{A.\arabic{equation}}
\label{app:BH_average}

In analogy with the EM case, to compute the average we will use a tool introduced by Brill and Hartle~\cite{Brill:1964zz}, to compute averages. Considering a tensor $T_{\mu\nu}(x)$, we defined the Brill-Hartle average by using
\begin{equation}
    \langle T_{\mu\nu}(x)\rangle \equiv \int d^4 y \,  f(x,y)\, g_\mu^{\tilde{\alpha}}(x,y)\, g_\nu^{\tilde{\beta}}(x,y)\, T_{\tilde{\alpha}\tilde{\beta}}(y)\, ,
\end{equation}
where $f(x,y)$ is a smooth weighting function which falls to zero when $\lambda \ll |\vec{x}-\vec{y}| \ll \mathcal{R}$, where $\mathcal{R}$ is the typical scale over which the background varies and $g_\mu^{\tilde{\alpha}}$ is the bi-vector of geodesic parallel displacement. It transforms as covariant vector w.r.t. $x$ for the index $\mu$ and as contravariant vector w.r.t. $y$ for the index $\tilde{\alpha}$. By using the Gauss theorem on curved manifolds, it is possible to show that the total covariant derivatives of an object quadratic in the GWs is negligible. More specifically, if we defined $S^\rho_{\mu\nu}$ as an object quadratic in the GWs, the Brill-Hartle average of its four-divergence can be written as 
\begin{equation}
    \begin{split}
    \left \langle \mathcal{D}_\rho S_{\mu\nu}^\rho \right \rangle  =& \int_{V}d^4 y \,  f(x,y)\, g_\mu^{\tilde{\alpha}}(x,y)\, g_\nu^{\tilde{\beta}}(x,y)\, \mathcal{D}_{\tilde{\rho}} 			S_{\tilde{\alpha}\tilde{\beta}}^{\tilde{\rho}}(y) = \\
    =& \int_{\Sigma}d\Sigma_{\tilde{\rho}} \, \,  g_\mu^{\tilde{\alpha}}(x,y)\, g_\nu^{\tilde{\beta}}(x,y)\, \mathcal{D}_{\tilde{\rho}} S_{\tilde{\alpha}\tilde{\beta}}^{\tilde{\rho}}(y)- \int_{V}d^4 y \,  \mathcal{D}_{\tilde{\rho}}\left[ f(x,y)\, g_\mu^{\tilde{\alpha}}(x,y)\, g_\nu^{\tilde{\beta}}(x,y)\right]\,  S_{\tilde{\alpha}\tilde{\beta}}^{\tilde{\rho}}(y)\, .
    \end{split}
\end{equation}
The boundary term is zero, because the smoothing function $f$ goes to zero at infinity, while other terms are negligible, since they vary on scales $L$. Therefore, the Brill-Hartle average of a four-divergence is suppressed by a factor 
\begin{equation}
    \left \langle \mathcal{D}_\rho S_{\mu\nu}^\rho \right \rangle \sim  \mathcal{D}_\rho S_{\mu\nu}^\rho \frac{\lambda}{L}\ll  \mathcal{D}_\rho S_{\mu\nu}^\rho\, .
\label{eq:negligible_total_derivative}
\end{equation}
In Appendix~\ref{app:Covariant derivatives of GWs}, we compute the four-divergences like Eq.~\eqref{eq:negligible_total_derivative} that can be neglected in the computation of the energy-momentum tensor of the CGWB.

\section{Covariant derivatives of GWs}
\label{app:Covariant derivatives of GWs}

\setcounter{equation}{0}
\renewcommand{\theequation}{B.\arabic{equation}}
\label{Appendice B}

The covariant derivatives of the radiative degrees of freedom of the metric are
\begin{equation}
\label{A.1}
    \begin{split}
        \mathcal{D}_0\left(\gamma_{ij}^{\rm GW}\right)=& a^2 h_{ij}^\prime\, , \\
        \mathcal{D}_k\left(\gamma^{\rm GW}_{ ij}\right)=& a^2\partial_k h_{ij}\, , \\
        \mathcal{D}_0\left(\gamma^{\rm GW\, ij}\right)=&\frac{(1+4\Phi)h^{ij\, \prime}-H^{k(i}h^{j)\, \prime}_k}{a^2}\, , \\
        \mathcal{D}_k\left(\gamma^{\rm GW\, ij}\right)=& \frac{(1+4\Phi)\partial_k h^{ij}-H^{l(i}\partial_kh^{j)}_l}{a^2} \, ,
    \end{split}
\end{equation}
where the round parentheses denote symmetrized indices. In these expressions we have neglected spatial derivatives of $\Phi$, $\Psi$, $H_{ij}$, which are subdominant on superhorizon scales. In addition, we neglect any friction term $\mathcal{H} h_{ij}$ or terms of the form $\Psi^\prime h_{ij}$, which are supposed to be suppressed w.r.t. terms containing the derivatives of $h_{ij}$, which depend on $q$. 

In the shortwave approximation, the tensors of the form $ \mathcal{D}_{\rho}S^{\rho}_{\mu\nu}\,$, where $S^\rho_{\mu\nu}$ is quadratic in $\gamma^{\rm GW}_{ij}$, can be dropped under averages ~\cite{Isaacson:1967sln}. We compute the covariant derivatives of the following combination at third order in perturbations 
\begin{equation}
    \begin{split}
    \label{eq:cov_derivatives_scalars}
    \mathcal{T}\equiv &\mathcal{D}_0\left(\gamma^{\rm GW\, ij}\mathcal{D}_0 \gamma_{ij}^{\rm GW}\right) =\\
    =&h^{ij\, \prime}h_{ij}^\prime(1+4\Phi)+h^{ij}h_{ij}^{\prime\prime}(1+4\Phi)-H^{k(i}h^{j)\, \prime}_kh_{ij}^\prime-H^{k(i}h^{j)}_k h_{ij}^{\prime\prime}\, , \\
    \mathcal{S}_{kl}\equiv &\mathcal{D}_k \left(\gamma^{\rm GW\, ij} \,  \mathcal{D}_l\gamma_{ij}^{\rm GW}\right)  \\
    \mathcal{S}\equiv& \delta^{m k}\mathcal{S}_{mk}=  (1+4\Phi)\left(\partial^k h^{ij}\partial_k h_{ij}+h^{ij}\nabla^2 h_{ij}\right)-H^{l(i}\left(\partial^k h^{j)}_l\partial_k h_{ij}+h^{j)}_l\nabla^2 h_{ij}\right) \, , \\
    \mathcal{M}\equiv &\mathcal{D}^\rho\left(\gamma^{\rm GW\, jk}\mathcal{D}_k \gamma_{j\rho}^{\rm GW}\right) = \\
    = & \frac{(1+6\Phi)\partial^i h^{jk} \partial_k h_{ij}-H^{s(k}\partial^i h^{j)}_s\partial_k h_{ij}-H^{im}(\partial_m h^{jk}\partial_k h_{ij}+h^{jk}\partial_m\partial_k h_{ij})}{a^2} \, .
    \end{split}
\end{equation}
Then the four-divergence can be written as
\begin{equation}
\begin{split}
    \mathcal{G}\equiv& \mathcal{D}^\mu\left(\gamma^{\rm GW\, ij}\mathcal{D}_\mu \gamma^{\rm GW}_{ij}\right) = \\
    = & -h^{ij\, \prime}h_{ij}^\prime(1-2\Psi+4\Phi)+(1+6\Phi)\partial^k h^{ij}\partial_k h_{ij} \\
    &+\mathcal{E}_{ij}\left[h^{ij}+2H^{li}h^j_l\right]+H^{l(i}h^{j)\, \prime}_l h_{ij}^\prime-H^{l(i}\partial_k h^{j)}_l \partial^k h_{ij}-H^{kl}\partial_k h^{ij}\partial_l h_{ij} \, ,
    \label{eq:four_divergence}
\end{split}
\end{equation}
with $\mathcal{E}_{ij}$ the perturbed equation of motion of the GWs introduced in Eq.~\eqref{eq:perturbed_eom_GWs}. We also exploit the following total derivatives at second order in perturbations 
\begin{equation}
    \begin{split}
        \label{eq:cov_derivatives_tensors}
        \mathcal{A}^{(0)}_{ij} \equiv &  \mathcal{D}_\rho\left(\gamma^{\rm GW\, \rho k}\mathcal{D}_j \gamma_{ik}^{\rm GW}\right) = h^{lk}\partial_l\partial_j h_{ik} \, , \\
        \mathcal{C}^{(0)}_{ij} \equiv& \mathcal{D}_\rho\left(\delta^\rho_i \gamma^{{\rm GW}\, kl}\mathcal{D}_j\gamma^{\rm GW}_{kl}\right) = \partial_i h^{kl}\partial_j h_{kl}+h^{kl}\partial_i\partial_j h_{kl} \, , \\
        \mathcal{D}_{ij}^{(0)} \equiv & \mathcal{D}^\rho\left(\delta^l_\rho \mathcal{D}_j \gamma^{\rm GW}_{kl} \gamma_i^{{\rm GW}\,k}\right) = \partial_j h_{kl} \partial^l h_i^k \, , \\
        \mathcal{F}_{ij}^{(0)} \equiv & \mathcal{D}_\rho\left(\gamma^{\rm GW}_{jl}\mathcal{D}^l \gamma_i^{\rm GW\, \rho}\right)= \partial_k h_{jl}\partial^l h_i^k \, , \\
        \mathcal{P}^{(0)}_{ij} \equiv &  \mathcal{D}^l\left(\gamma_i^{{\rm GW}\, k} \mathcal{D}_l \gamma^{\rm GW}_{jk}\right)= \partial^l h_i^k\partial_l h_{jk}+h_i^k\nabla^2 h_{jk} \, .
    \end{split}
\end{equation}

\section{The Einstein tensor at second order}
\setcounter{equation}{0}
\renewcommand{\theequation}{C.\arabic{equation}}
\label{App:The Einstein tensor at second order}
In order to be sure that the computation of the Einstein tensor at first order in the large-scale perturbations is correct, we check the consistency of our
results with the computation of the Einstein tensor at zero order in the large-scale perturbations and quadratic in $h_{ij}$ associated to the average energy density of GWs given in Eq. \eqref{eq:rho_GW_tot} that can be found in the literature. To perform the computation of the Einstein tensor we use xPand~\cite{Pitrou:2013hga}. The output of the code for the Einstein tensor at second-order is
\begin{equation}
    \begin{split}
        a^2 G^{0\, \, (2)}_0 = & \frac{1}{8}h^{ij\, \prime} h_{ij}^\prime+\mathcal{H} h^{ij}h_{ij}^\prime-\frac{1}{2} h^{ij} \nabla^2 h_{ij}+\frac{1}{4}\partial^i h^{jk}\partial_k h_{ji}-\frac{3}{8}\partial^k h^{ij}\partial_k h_{ij}\, .
        \label{eq:ET_snd_order} 
    \end{split}
\end{equation}
The shortwave approximation allows to simplify the expression we get for the Einstein tensor. We can rewrite the Einstein tensor at second order in $h_{ij}$ and zero order in the large-scale perturbations as
\begin{equation}
    \begin{split}
        a^2 G^{0\, \, (2)}_0 = & \frac{1}{4} h^{ij\,\prime} h_{ij}^\prime +\frac{1}{8}h^{ij}\mathcal{E}_{ij}^{(0)}+\frac{1}{8}\mathcal{T}^{(0)}-\frac{1}{4}\mathcal{M}^{(0)}+\frac{3}{8}\mathcal{S}^{(0)} \, ,  
        \label{eq:emt_second_order}
    \end{split}
\end{equation}
where we neglected the friction term proportional to $\mathcal{H}$ and the perturbed equation of motion defined in Eq.~\eqref{eq:eom_GWs_unperturbed}.  Moreover, we can exploit the fact that the averaged total covariant derivatives defined in Appendix \ref{app:BH_average} are equal to zero. This leads to 
\begin{equation}
    \begin{split}
        a^2 \langle G^{0\, \, (2)}_0 \rangle= & \frac{1}{4} \langle h^{ij\,\prime} h_{ij}^\prime \rangle  \, ,  
        \label{eq:emt_second_order}
    \end{split}
\end{equation}
and we get the standard expression for the average energy density of GWs
\begin{equation}
        \bar{\rho}_{GW} =   - T^{0\, \, }_0  = \frac{1}{32\pi G a^2 \left(\eta\right) }\langle h_{ij}' h^{ij\, \prime}\rangle  \, .
\end{equation}

\bibliographystyle{ieeetr}
\bibliography{bibliography.bib}

\end{document}